\documentstyle[prl,aps,multicol,epsf]{revtex}

\begin{document}
\draft
\tighten

\title{Emergent $Z_2$ gauge symmetry and spin-charge separation in one
dimensional physics.}
\author{H.V. Kruis, Z. Nussinov and J. Zaanen}
\address{Instituut-Lorentz for Theoretical Physics, Leiden University\\
P.O.B. 9506, 2300 RA Leiden, The Netherlands}
\date{\today ; E-mail:hvkruis@lorentz.leidenuniv.nl;
zohar@lorentz.leidenuniv.nl; jan@lorentz.leidenuniv.nl
}
\maketitle

\begin{abstract}
We demonstrate that in the scaling limit the phenomenon of spin-charge 
separation
as encountered in Luttinger liquids defined on the lattice can be associated
with the emergence of local Ising symmetry. This $Z_2$ gauge
field is of geometrical nature (sublattice parity) and the Luttinger liquid 
corresponds with the critical point where the symmetry is on the verge of
becoming local. We argue that in a true one dimensional superconductor the
local symmetry will be fully realized as long as the geometrical structure
associated with spin-charge separation is maintained.  
 \end{abstract}

\begin{multicols}{2}
\narrowtext

It is quite well understood how to break  symmetry, but a different 
matter is how to {\em make} symmetry. With the exception of
electromagnetic $U(1)$, the manifest symmetries governing 
condensed matter physics are of a global nature. Nevertheless, 
the idea that gauge theory is behind the peculiarities of high $T_c$
superconductivity is quite influential, especially so in the form
of the slave theories of spin-charge separation\cite{slave,sentfish}. 
Here we will demonstrate that behind the phenomenon of spin-charge 
separation in one spatial dimension a dynamically generated 
local symmetry of the 
Ising ($Z_2$) kind is hidden. This gauge principle is different from those 
envisaged in the slave theories\cite{weng}. It is rooted in geometry:  
the gauge fields parametrize the fluctuations of the geometric property 
bipartiteness. A lattice is called bipartite when it can be subdivided 
in two sublattices $A$ and $B$, such that all sites on the $A$ sublattice
are neighbored by $B$ sublattice sites and vice versa. This division can be
done in two ways ($\cdots - A - B - A - B \cdots$ and $\cdots - B - A - B - A
\cdots$) defining a $Z_2$ valued quantity `sublattice parity', $p = \pm 1$. 
Bethe-Ansatz results show that in the scaling limit sublattice parity acquires
the status of separate dynamical degree of freedom. The quanta of electrical 
charge are bound to flips in the sublattice parity which in turn 
alter the space seen by the spin system.
Sublattice parity turns into a $Z_2$ gauge field when the quantization 
of charge is destroyed by true superconducting long range order, while
the Luttinger liquid is right at the quantum phase transition where this 
local $Z_2$ symmetry emerges.   

The above claims are based on a re-analysis of exact Bethe-Ansatz 
results for the one dimensional Hubbard model\cite{liebwu}.
Some time ago, Ogata and Shiba\cite{ogashi} discovered a simple but most peculiar 
property of the Bethe Ansatz wavefunction in the special case that $U$ 
is very large ($U \rightarrow \infty$ limit). They showed that 
the wave function factorizes in a charge part $\psi_{SF}$, depending 
on where the electrons are, and a spin part $\psi_{H}$
which is merely depending on the way the spins are distributed,
\begin{equation}
\psi(\{ x_{i} \}_{i=1}^{N}; \{ y_{j} \}_{j=1}^{N/2}) = \psi_{SF}
(\{x_{i} \}_{i=1}^{N})~ \psi_H (\{ y_{j} \}_{j=1}^{N/2}).
\label{ogatashiba}
\end{equation}
The charge part $\psi_{SF}$ is nothing else than the wave-function of a
non-interacting spinless-fermion system where the coordinates $x_i$ refer
to the actual positions of the electrons/spinless fermions. The spin part
$\psi_H$ is identical to the wavefunction of a chain of Heisenberg spins
interacting via a nearest-neighbor antiferromagnetic exchange. Although
this factorization property gives a precise meaning to the notion of  spin-charge
separation, charge and spin are actually not quite independent from each
other. In $\psi_H$ only the positions
of the up spins are needed and these correspond with the coordinates $y_i$. 
The surprise is now that the coordinates $y_i$ do not refer the original
Hubbard chain, but instead to a {\em new space}: a
lattice with sites at coordinates $x_{1},  x_{2}, ..., x_{N}$ given by  
the positions of the charges in a configuration with 
amplitude $\psi_{SF}$. 

The above gives away that the
quantum dynamics of interacting electrons generates a {\em geometrical 
structure} analogous to the fabric of general relativity.
Let us visualize this for a representative example (Fig. 1). Consider
 $N$ electrons
on a chain with $L$ sites under the condition that $N < L$ such that the charge
configurations can be specified by the locations of the holes.
A charge configuration in the full Hubbard chain (`external space'), 
has an amplitude  $\psi_{SF}$ in the wavefunction with the 
coordinates of the dots corresponding 
with the $x_i$'s. The spin system sees a different `internal space' 
obtained from the full space  by removing the holes {\em together with the
sites where the holes are located}, substituting the hole and its site with an
antiferromagnetic exchange between the sites neighboring the hole (the
`squeezed space'\cite{ogashi}).  

As in general relativity, physics is derived from relations
between the different reference frames,
but the geometry involved in the Ogata-Shiba case is obviously much simpler than
the geometry of fundamental space-time. This simplification makes it possible
to parametrize matters in terms of a simple gauge theory.
In which regards are the full chain and the squeezed chain different? 
The squeezed chain is obviously shorter than the full chain and this is
a simple dilation: a distance $x$ measured in the full chain becomes a distance
$\rho x$ in the squeezed chain ($\rho = N / L$, the electron density)
 when $x \gg 1$,
the lattice constant. The other aspect is also simple, but less trivial in its
consequences. The spin system is a quantum-antiferromagnet and it is as such
sensitive to the geometrical property of bipartiteness. Consider what happens with
the sublattice parity, as introduced in the first paragraph. For the
Heisenberg spin chain, as the one in squeezed space, a redefinition of $p = 1
\leftrightarrow -1$ does not carry any consequence (`pure gauge').
However, sublattice parity becomes alive 
in the mapping of squeezed space into full space (Fig. 1). `Fix the gauge'
in squeezed space by choosing a particular sublattice parity, and consider
what happens when it is unsqueezed. The holes are inserted, and because every  
hole is attached to one site,  every time a hole is passed the
sublattice parity flips.  

The above is true for every instantaneous charge configuration. However, the
ground state is a superposition of many of these configurations. In 1+1D
charge cannot break translational invariance and  since this fluctuating charge is
`attached' to the sublattice parity flips, the full space which is observable
by external observers (experimentalists) should be considered as a fluctuating 
geometry. However, this is a very simple fluctuating geometry because all what is 
fluctuating is the property of bipartiteness. This
geometrical fluctuation can be parametrized in terms of a field theory 
controlled by the simplest of all local symmetries: $Z_2$ gauge theory\cite{kogut}.

\begin{figure}[h]
\hspace{0.0 \hsize}
\epsfxsize=0.9\hsize
\epsffile{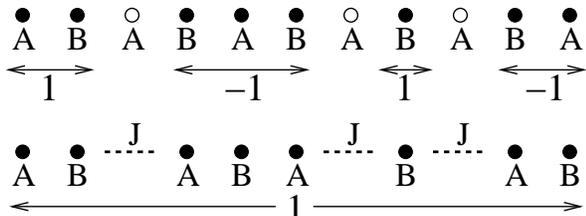}
\caption{The emergence of the $Z_2$ `sublattice parity field' in the
geometrical `squeezing' operation. The open circles refer to some
configuration of holes. The black dots define the embedding space of
the spin system in both the full- (upper) and squeezed (lower) lattice.
The unsqueezing operation can be parametrized in terms of binding of
the electric charge quantum to flips in the $Z_2$ valued sublattice
parity field.} 
\label{f1}
\end{figure}

Let us start out on the geometrical level and ask the question if a correlation
function can be defined acting on the full Hubbard chain which can measure
the `true' spin correlations associated with squeezed space. Since all what
matters is sublattice parity this can be achieved by simply multiplying
the spin operator by a factor $-1$ every time a hole is passed, thereby 
removing the sublattice parity flips from the spin correlations. Define
staggered magnetization as $\vec{M} (x_i) = (-1)^{x_i} \vec{S} (x_i)$ where
$\vec{S}$ is the spin operator ($S_z = n_{\uparrow} - n_{\downarrow},
S^{+} = c^{\dagger}_{\uparrow} c_{\downarrow}$) and the charge operator 
$n_{x_i} = n_{x_i \uparrow} + n_{x_i \downarrow}$ taking the values 0,
1 and 2 for an empty-, singly- and doubly occupied site, respectively. The
correlation function we are looking for is \cite{zavsa},
\begin{eqnarray}
O_{top} (| x_i - x_j |) = \langle M^{z} (x_i) \Pi_{x_l = x_i+1}^{x_j-1}
e^{i \pi ( 1 - n_{x_l} ) } M^{z} (x_j)  \rangle.
\label{otop}
\end{eqnarray}
The operator $\exp ( i \pi ( 1 - n (x_l))$ takes the value $+1$ for a singly
occupied spin/squeezed space site while it is $-1$ for a charge (hole, or
doubly occupied) site. By multiplying these values on the interval 
$x_i \le x_l \le x_j$ all the minus signs associated with the sublattice
parity flips are removed from the spin correlations.

Although the `string' operator $\sim \Pi \exp ( i \pi (1 - n))$ is non-local
it can be evaluated straightforwardly using the techniques introduced by
Parola and Sorella\cite{parsor}. It is easily shown that
\begin{eqnarray}
\langle S^{z}(0)\, &(-1)&^{\sum_{x_j=1}^{x_i-1} \left[ 1- n(x_j) \right] } 
\, S^{z}(x_i) \, \rangle 
\nonumber \\ 
&=& \sum_{x_j=2}^{x_i+1} P_{SF}^{x_i} (x_j)
(-1)^{x_i+1-x_j} O_{Heis.} (x_j-1)
\label{parsor}
\end{eqnarray}
where $O_{Heis.}$ is the spin correlator of the Heisenberg chain,
while $P^{x_i}_{SF} (x_j) = \langle \,  n(0) \,  n(x_i) \,
\delta (\sum_{x_l=0}^{x_i} n(x_l) -x_j) \, \rangle _{SF}$ is the probability of
 finding $j$ spinless fermions in the interval $\left[0,x_i \right]$. This
factor causes the additional decay of the spin correlations due to the
charge fluctuations in the standard spin correlator. However, it is easily 
shown that it is precisely compensated in $O_{top}$ by the factor 
$(-1)^{x_i+1-x_j}$ coming from the string operator and we find the result, 
asymptotically exact for large distances $x$,
\begin{equation}
O_{top} ( x ) = {\rho \over x } \ln^{1/2} ( \rho x ).
\label{hubtop}
\end{equation}
Does this reveal the spin-only problem in squeezed space? We have to correct
for the fact that the density of staggered spin in full space is reduced by a factor
$\rho$ as compared to squeezed space while in addition distances are  measured
by $ x / \rho$ instead of $x$. After rescaling,
$ O_{top}^{scaled} (x) = {1 \over x} \ln^{1/2} (x)$
which is indeed the behavior of the staggered spin correlation function
of a Heisenberg spin chain at large distances.

Let us now consider the well known\cite{parsor} asymptotic behavior of the 
`normal' staggered spin correlations for the Hubbard chain\cite{zastrgas}, 
\begin{eqnarray}
O_{\vec{M}} (|x_i - x_j|) &  = & \langle | M^{z} (x_i) M^{z}(x_j) | \rangle
\label{OM} \\ 
    & \sim & { { \cos ( \varepsilon (\rho) x) } \over x^{1 + K_{\rho}} }.
\label{OMHub}
\end{eqnarray}
What matters  is that the algebraic decay is more rapid than for the
correlator  Eq. (\ref{hubtop}). The charge-stiffness $K_{\rho}$ is associated 
with the decay of the charge correlations, $\langle | n (x_i) n (0) | \rangle
\sim  \cos ( 2 \varepsilon x)  / x^{K_{\rho}}$, and $K_{\rho} > 0$ for all densities
$\rho \neq 1$. Hence, the charge 
fluctuations influence  the spin correlations because the charge is attached 
to the sublattice parity flips, and they  do so by a simple multiplicative factor
$1 / x^{K_{\rho}}$. 

Is the above an accident of the strongly coupled
case? Let us consider the case where the interactions are completely vanishing:
the spinfull fermion gas on the lattice. It is straightforward to show that in
this limit
$\langle \, S^z(x_i) \, 
(-1)^{\sum_{x_j=1}^{x_i-1} \left[ 1- n(x_j) \right] } \, S^z(0) \, \rangle =
 \frac{ (-1)^{x_i-1}  }{4} \left[ 
D(x_i-2) D(x_i) -D(x_i-1)^2
 \right]$
where $D(x-i) \equiv \langle \, (-1)^{\sum_{x_j=0}^{x_i} n(x_j) } \rangle $. 
This determinant
can be evaluated numerically and we find that $O_{top} (x) \sim 1/ x$. As
$O_{\vec{M}} \sim 1 / x^{K_{\rho} + K_{\sigma}} \sim 1 / x^2$, while $K_{\rho}
= 1$ in the non-interacting case, it follows that there is  
no difference in the asymptotics between the $U \rightarrow
\infty$ and the $U = 0$ case and by continuity it has to be that
the squeeze is true at least for all $ U \ge 0$. The exact limiting
results ($U=0, \to \infty$) are further fortified by bosonization
for arbitrary $\rho$
which once again yields $O_{top} \sim x^{- K_{s}}$
\cite{bosonization}. It is remarkable that free fermions 
code for a phenomenon as involved as the Ogata-Shiba squeeze!

We have now gathered the information needed to make the case that a local Ising
symmetry is at work.  The key is 
that the non-local charge string $\sim \Pi \exp ( i \pi ( 1 - n) )$ can be 
rigorously identified with the Wilson line of a $Z_2$ gauge theory\cite{kogut}, 
with the logical implication that the different decay rates of 
$O_{top}$ and $O_{\vec{M}}$
are manifestations of the local symmetry. The charge sector can be parametrized
for large $U$ in terms of spinless fermions, and these can in turn be expressed
in terms of Pauli matrices $\sigma^{\alpha}$ using a Jordan-Wigner transformation,
\begin{eqnarray}
\sigma_{x_i}^3 & = & 1 - 2 n_{x_i} \nonumber \\
\sigma_{x_i}^+ & = & [ \Pi_{x_j < x_i} \sigma^3_{x_j} ] f_{x_i} \nonumber \\
\sigma_{x_i}^- & = & [ \Pi_{x_j < x_i} \sigma^3_{x_j} ] f^{\dagger}_{x_i}
\label{jorwig}
\end{eqnarray}
where $f^{\dagger}_{x_i}$ creates a spinless fermion, while their number $n_{x_i} =
f^{\dagger}_{x_i} f_{x_i}$ is identical to the electron charge operator $n$ defined
in the above. Hence, by operator identity we can write the vertex operator as,
\begin{equation}
e^{i\pi ( 1 - n_{x_i} ) } =  e^{ i (\pi/2) ( 1 + \sigma_{x_i}^3)}
=  \sigma^3_{x_i}.
\label{vertid}
\end{equation}
As charge and spin are 
decoupled, we may shift the charge degrees of 
freedom to the lattice formed by the 
midpoints of the bonds between the sites of 
the original lattice (the bond-, or 
dual lattice,  with coordinates $x_i,x_{i+1}$) while 
concurrently retaining the spin degrees of 
freedom on the original lattice. After these operations, our
correlator Eq. (\ref{hubtop}) becomes,
\begin{equation}
O_{top} (|x_i - x_j|) = \langle | M^{z}_{x_{i}} [ \Pi_{<x_l,x_{l'}> \in \Gamma} 
\sigma^3_{x_l,x_{l'}} ]
M^{z}_{x_j} | \rangle.
\label{gaugeinv}
\end{equation}
This is nothing else than the gauge invariant correlation function of a theory
containing a $SU(2)$ `matter' field $\vec{M}$ coupled to an Ising gauge field.
The factor $\Pi \sigma^3$ corresponds with the Wilson-line of the gauge
theory ($\Gamma$ is a line connecting $x_i$ and $x_j$), which has to be inserted
to keep the matter-field correlator gauge invariant. By the same token, the
`normal' spin correlator $O_{\vec{M}}$ is not gauge invariant, because the
Wilson line is missing, and our observation that $O_{\vec{M}}$ is more rapidly
decaying than $O_{top}$ signals the presence of  local $Z_2$ symmetry.

Let us explain this in some more detail\cite{kogut}. 
Imagine that an effective, long wavelength
theory is realized with a Hamiltonian invariant under the following $Z_2$
gauge transformation ($\sigma_{x_i,x_{i+1}}$ is the value of the $Z_2$ variable
on bond ${x_i,x_{i+1}}$),
\begin{eqnarray}
\sigma_{x_i, x_{i+1}} & \rightarrow & \eta_{x_i} 
\sigma_{x_i, x_{i+1}} \eta_{x_{i+1}} \nonumber \\
\vec{M}_{x_i} & \rightarrow & \eta_{x_i} \vec{M}_{x_i}
\label{gaugetr}
\end{eqnarray}
with arbitrary $\eta_{x_i} = \pm 1$. Given this symmetry, some very
general statements follow  regarding the behavior of correlation functions. First consider the spin 
correlator $O_{\vec{M}} (|x_i - x_j|) = \langle M^{z}(x_i) M^{z} (x_j) \rangle$.
Since $O_{\vec{M}} \rightarrow -O_{\vec{M}}$ under the transformation 
Eq. (\ref{gaugetr}) at either $x_i$ or $x_j$ it has to vanish: it is not 
gauge invariant.  Since the gauge symmetry is  emergent an energy scale should exist
below which it becomes active. An energy scale implies a length scale 
$l_{gauge}$ and therefore  $O_{\vec{M}} (x) \sim \exp ( -x / l_{gauge} )$. 
The unique way to construct gauge invariant 
correlation functions is by inserting a closing Wilson line,
$\Pi_{\Gamma} \sigma^3$, as 
done in $O_{top}$ (Eq. (\ref{gaugeinv})).  
The minus sign introduced by the gauge 
transformation of e.g.  $\vec{M} (x_i)$ is compensated
by the minus sign introduced by the simultaneous transformation of the bond variable
on the bond leaving site $x_i$ in the direction of site $x_j$. However, a single 
$\sigma^3$ is not gauge invariant and to maintain overall gauge invariance
the string $\Pi \sigma^3$, ending at $x_{j}$, needs to be 
inserted. The behavior
of $O_{top}$ will depend on the details of the dynamics of the gauge theory.
However, one statement is always valid: $O_{top}$ will decay more slowly
then $O_{\vec{M}}$ when the gauge invariance is present, and this is what we  
found to be the case in the Luttinger liquid. 

Nonetheless, there is a complication. As we argued, an unavoidable
consequence of a full realization of the $Z_2$ gauge symmetry is 
that correlation function violating the gauge invariance 
 should decay exponentially faster than
the gauge invariant ones, and in the Luttinger liquid
we found  that this difference is a mere
algebraic factor. Algebraic behavior of correlation functions is associated with 
yet another symmetry: scale invariance.  Since $O_{top}$
shows a slower algebraic decay than $O_{\vec{M}}$  it 
has to be that the Luttinger liquid is right at the critical point where the $Z_2$
symmetry turns on, which is associated with the divergence of the gauge
length: $\l_{gauge} \rightarrow \infty$.

This is less mysterious than it might sound. It is clarifying to view this from
the geometrical (squeezed space) perspective. The $Z_2$ local invariance just
means that it is undetermined how many sublattice parity flips occur per unit length.
Since these flips are attached to the charge quanta $Z_2$ invariance is established
when charge is a fluctuating quantity. $Z_2$ turns into a global invariance in a 
state where charge is locally conserved: the Wigner (or `holon') crystal -- it is
easily checked that $O_{top}$ and $O_{M}$ behave identically at long distances when
the holes are assumed to be localized. In one dimension the tendency to crystallize
is universal\cite{zastrgas}. However, this order always turns into algebraic order 
due to the admixing of the zero-modes and algebraic order is a form of criticality.
This turns into criticality of the gauge sector due to the
charge-sublattice parity binding.

Hence, the system can be driven away from the phase transition to the phase 
obeying global $Z_2$ by breaking explicitly translational invariance. 
Is it possible to drive it in the other direction,
into a phase where the local $Z_2$ invariance is truly realized? The answer is
simple: add a field stabilizing true superconducting long range order in the charge
sector. The charge fluctuations which are fundamental to a system with
phase order are of precisely the right kind to protect the local $Z_2$ symmetry.
By adiabatic continuity, a system showing true superconducting order
can be thought of as composed of lattice  bosons $b^{\dagger}(x_i)$ carrying
a charge q. In addition, it is a requirement that this charged boson is attached to
the sublattice parity flip\cite{ladders}. For long wavelength
purposes, the charge part
of the ground state wave-function of  
the phase-ordered state may be written as
\begin{equation}
| \psi_{q} \rangle = \Pi_{x_i} [ u + v b^{\dagger} (x_i) ] | vac \rangle.
\label{bossup}
\end{equation}
(where, in the low density limit, $|u| \gg |v|$)
such that the superconducting order parameter $\Psi = u^{*} v$. It follows that
($l_{\Gamma}$ is the length of the Wilson line),
\begin{eqnarray}
\langle \psi_{q} | \Pi_{\Gamma} e^{i \pi b^{\dagger} (x_i) b (x_i) }
| \psi_{q} \rangle & = & ( |u|^2 - |v|^2 )^{l_{\Gamma}} \nonumber \\
                   & = & e^{ - { { l_{\Gamma} } \over {\l_{gauge} } } }
\label{supwil}  
\end{eqnarray}
where $l_{gauge} = -1 / \ln ( |u|^2 - |v|^2 )$. This in turn implies,
\begin{equation}
O_{\vec{M}} (| x_i - x_j|) = e^{ - { {|x_i - x_j|} \over {l_{gauge} } } } 
O_{top} (| x_i - x_j |) 
\label{eureka}
\end{equation}
proving that a full local $Z_2$ invariance is generated. The key is that
the spontaneous breaking of charge $U(1)$ symmetry is the same as the spontaneous
making of local $Z_2$ symmetry, gauging the spin sector!

Besides the clarification of the symmetries underlying one dimensional
physics, does this new symmetry principle buy us new physics in
1+1D in the sense of `new states of matter'? This is disappointing,
for no other reason than that gauge theories are not particularly
interesting in 1+1D
\cite{kogut}. The reason is that in one spatial dimension the
gauge field can only carry kinetic energy: using Hamiltonian language
the only gauge invariant operator 
available to describe the dynamics of the gauge field is $\sigma^1_{x_i,x_{i+1}}
= ( \sigma^+_{x_i,x_{i+1}} + \sigma^-_{x_i,x_{i+1}} ) / 2 $.
Hence, the ground state is always confining (eigenstate of $\sum \sigma^1$),
giving rise to exponential decay of all spin-related correlation
functions. This can be directly checked for our superconductor. In analogy
with Eq. (\ref{jorwig}), $\sigma^1 \sim b^{\dagger} + b$ which acquires a
vacuum amplitude because $\langle \psi_q | b^{\dagger} + b | \psi_q \rangle$
is non-zero in a superconductor.  
The situation is more interesting in two (and higher) spatial dimensions, because
the gauge field can now carry also a potential energy.  A variety
of distinguishable phases becomes possible\cite{sentfish,fradkinshankar,toner}.
Elsewhere we will analyze the 2+1 dimensional generalization of the above,
clarifying  the meaning of  the cuprate stripes and suggesting a novel way of 
viewing the high $T_c$ enigma.   

{\em Acknowledgments.} 
We acknowledge stimulating discussions with S.A. Kivelson,
S. Sachdev, F. Wilczek,
N. Nagaosa and T.K. Ng. Financial support was provided by the Foundation
of Fundamental Research on Matter (FOM), which is sponsored by the
Netherlands Organization for the Advancement of Pure Research (NWO).

\end{multicols}

\end{document}